\documentclass[12pt]{article}
\usepackage[english]{babel}
\usepackage[paper=letterpaper,top=1in,bottom=1.7in,inner=1in,outer=1in]{geometry}
\usepackage{setspace}
\usepackage{tikz}
\usepackage{appendix}
\usepackage[makeroom]{cancel}
 \usepackage{multirow}
\usepackage{amssymb}
\usepackage{bm}
\usepackage{graphicx}
\usepackage[normalem]{ulem}
\usepackage[cmex10]{amsmath} 
\usepackage{amsthm}
\usepackage{url}
\usepackage[square,sort,comma,numbers]{natbib}
\usepackage{hyperref}
\usepackage{caption}
\usepackage{subcaption}
\usepackage[nofiglist,notablist]{endfloat}
\newcommand{\Ex}{\ensuremath{\mathsf{E}}}

\theoremstyle{plain}

\theoremstyle{definition}

\begin{document}

\title{Response adaptive designs for binary responses: how to offer patient benefit while being robust to time trends?}
\author{Sof\'ia S. Villar$^{1}$\footnote{Corresponding author: {\sf{e-mail: sofia.villar@mrc-bsu.cam.ac.uk}}, Phone: +44 (0)1223 330385, Fax: +44 (0)1223 330365 $^{1}$ MRC Biostatistics Unit, Cambridge Institute of Public Health,
Forvie Site, Robinson Way, Cambridge Biomedical Campus, Cambridge CB2 0SR
United Kingdom. $^{2}$ MRC Integrative Epidemiology Unit, University of Bristol Oakfield House Bristol BS8 2BN},
Jack Bowden$^{2}$
and
James Wason$^{1}$}
\date{}
\maketitle
\begin{abstract} 

 Response-adaptive randomisation (RAR) can considerably improve the chances of a successful treatment outcome for patients in a clinical trial by skewing the allocation probability towards better performing treatments as data accumulates. There is considerable interest in using RAR designs in drug development for rare diseases, where  traditional designs are  not feasible or ethically objectionable. In this paper we discuss and address a major criticism of RAR: the undesirable type I error inflation due to unknown time trends in the trial. Time trends can appear because of changes in the characteristics of recruited patients -  so-called ‘\emph{patient drift}’.  \emph{Patient drift} is a realistic concern for  clinical trials in rare diseases because these typically recruit patients over a very long period of time. We compute by simulations how large the type I error inflation is as a function of the time trend magnitude in order to determine in which contexts a potentially costly correction is actually necessary. We then assess the ability of different correction methods to preserve type I error in this context and their performance in terms of other operating characteristics, including patient benefit and power. We make recommendations of which correction methods are most suitable in the rare disease context for several RAR rules, differentiating between the two-armed and the multi-armed case. We further propose a RAR design for multi-armed clinical trials, which is computationally cheap and robust to several time trends considered.
\end{abstract}







\section{Introduction}\label{sec:intro}
Randomised controlled trials (RCTs) are considered the gold standard approach to learn 
   about the relative efficacy of competing treatment options for evidence based patient care. The information provided by a RCT can subsequently  be used to better treat future populations. 
   Traditionally, patients are allocated with a fixed and equal probability to either an experimental treatment or standard therapy arm.
   We will refer to  RCTs implemented in this way as incorporating complete randomization (CR). 
However, there is generally a conflict between the individual benefit of patients in the trial and the collective benefit of future patients. CR, by definition, does not provide the flexibility to alter the allocation ratios to each arm, even if information emerges that breaks the initial trial equipoise.

 Response adaptive randomisation (RAR)
 offers a way of simultaneously learning about treatment efficacy while also  benefiting patients  inside   the trial.  It achieves this by skewing allocation to a better performing treatment, if it exists, as data is accrued. When RAR rules are used in a  multi-armed
trial, they also increase the probability
of finding a successful 
 treatment 
and 
speed up the process of doing so \citep{meurer2012adaptive, Wason}.

However, RAR is still infrequently used in practice. One of the most prominent recent arguments against its use is the concern that the false positive error rate (or Type I error rate) may not be controlled at the nominal level \citep{Thall2015}. This can  be the case if the distribution of patient outcomes changes over time independently of any treatment effect,   and the traditional methods of analysis are used \citep{simon2011}. One such example is when the underlying prognosis of patients recruited in the early stages of a trial differs from those recruited in the latter stages. This is often referred to as `patient drift'. \citep{karrison2003group} investigate the type I error inflation induced by various RAR rules implemented within a two-armed group sequential design with a binary outcome in which, depending on the observed value of the corresponding $z$-statistics, the next group of patients is allocated in one of four possible fixed ratios $R(z)$. They show that  if all success rates increase by $0.12$ over the course of a study with three interim analysis, the type I error rate achieved by a group sequential  design is
`unacceptably high', with the inflation being worst for the most aggressive RAR rules 

Time trends are more likely to occur in studies that have a long duration. 
Consider for example, the Lung Cancer Elimination (BATTLE)-1 phase II trial which recruited patients for 3 years (2006-2009). It was found that more smokers and patients who had previously received the control treatment enrolled in the latter part of the study compared to the beginning of the study \citep{liu2015overview}.
Trials that last more than 3 years will often be required 
for rare diseases because of the recruitment challenge. It is also exactly for this case where  the use of RAR can be  most desirable as the trial patients represent a higher proportion of the total patient population.
There have been several papers comparing different classes of RAR rules under various perspectives \citep[see e.g.][]{ivanova2000comparison,gu2010simulation,flournoy2013graphical,biswas2016response}
however, there has been very little work in the literature assessing the impact of time trends on different RAR rules and on how it can be best addressed 
for each of these rules
. 

In the paper by \cite{Thall2015} several criticisms to the use of RAR are given, including the type I error inflation under time trends. However, the paper only illustrates the existence of these drawbacks for a special class of RAR (based on regular updates of posterior probabilities). 
In the trial context investigated by \citep{karrison2003group}
 an analysis stratified by trial stage  eliminates the type I error  inflation
induced by a simple upward trend of all the success rates. \citep{simon2011} considered broad RAR rules for the two-armed case and proposed a randomisation test to correct for type I error inflation caused by unknown time trends of any type.
In \citep{rosenberger2001covariate} a covariate-adjusted response adaptive mechanism for a two-armed trial that can take a specific time trend as a covariate is introduced.
In this paper we identify and address a number of unanswered questions which we describe below, including the study of the multi-armed case.

If one is considering designing a clinical trial using a particular RAR rule then a fundamental question to consider is how large  the temporal change in the trial data has to be to materially affect the results. 
In \autoref{Sec:2}  we address this question for a representative selection of RAR procedures.

If the possibility of a large drift occurring during the trial is a concern and a RAR scheme is being considered for designing such a trial, then
subsequent and related questions are: Do any `robust' hypothesis testing procedures exist that naturally preserve type I error in the presence of an unknown
time trend? Should these procedures be different for two-armed  and for multi-armed trials? 
Should they differ depending on the RAR rule in use? 
And, finally, what is their effect on statistical power?
\autoref{Sec:2} and \autoref{Sec:3} address these questions for different RAR
procedures. In \autoref{Sec:4} we consider whether time trends can be effectively detected and adjusted for in the analysis, and how extended modelling approaches  for modelling a time trend compare to 
model-free approaches in order to control for type I error.
In \autoref{Sec:5} some
conclusions and recommendations for addressing this specific concern are given.

\section{RAR rules, time trends and type I error rates}\label{Sec:2}

In this section we assess the impact of different time trend assumptions on the type I error rate of distinct RAR procedures.
We assume that patients are enrolled in the trial sequentially, in groups of equal
size $b$ over $J$ stages. We do not consider monitoring the trial for early stopping  and therefore the trial size is fixed and equal to  $T=b \times J$. We have omitted it the possibility of early stopping in this paper to isolate the effects of an unaccounted for time trend in a trial design using RAR as the only form of adaptation included.
Patients are initially allocated with an equal probability to  each  treatment
arm.  After the first interim analysis allocation probabilities will be updated based on data and according to different RAR rules. In a real trial, this initial CR start-up phase could be replaced by a restricted randomisation phase (e.g. a permuted block design) to minimise sample imbalances and improve the subsequent probabilities updates \citep{haines2015}.  For simplicity of presentation we consider a binary outcome variable $Y_{i,j,k}$ for patient $i$ allocated to treatment $k$ at stage $j$, (with $Y_{i,j,k}=1$  representing a success and $Y_{i,j,k}=0$  a failure) that is observed
relatively quickly after the allocation. 
An example might be whether a surgery is considered to have been successful. 

We consider a trial with $K\ge1$ experimental arms and a control arm and assume all patients in block $j$ are randomised to treatment $k$  with probability  $\pi_{j,k}$ (for  $j=1, \dots, J$  and $k=0,1,\dots, K$). For example, a traditional CR design will have $\pi_{j,k}=1/(K+1) \quad \forall j,k$. Patient treatment allocations are recorded by binary variables $a_{i,j,k}$ that take the value 1 when patient $i$ in block $j$ is allocated to treatment $k$ and $0$ otherwise. We assume that every patient in the trial can only receive one treatment and therefore we impose that $\sum_{k=0}^{K} a_{i,j,k}\le 1$ for all $i,j$. We will also assume that for every $j<J$ before making the treatment decisions for the $(j+1)^{th}$ block of patients the outcome information of the $j^{th}$ block of patients is fully available. We denote the control treatment by $k=0$. Updating the allocation probabilities after blocks of patients rather than after every patient makes the application of RAR rules more practical in real trials \citep{rosenberger1993use}.  

An appropriate test statistic is used to test the hypotheses that the outcome probability in each experimental
treatment is equal to that of the control. That is, if we let $Pr(Y_{i,j,k}=1|a_{i,j,k}=1)=p_k$, 
then we  consider the global null 
to be $H_{0,k}: p_0 = p_k$ for $k=1,\dots,K$.
Generally, any sensible test statistic will produce valid inferences if the outcome probability in each arm conditional on treatment remains constant over the course of the trial. If this is not the case, then the analysis may be subject to bias. To illustrate this 
%
we shall assume the following model for the outcome variable $Y$
\begin{equation}\label{model1}
\text{Logit}\left[Pr(Y_{i,j,k}=1|Z_{i,j}, a_{i,j,k}=1)\right] =
\begin{cases} 
  \beta_{0} +  \beta_{t}t_{j}+ \beta_{z}Z_{i,j} & k=0 \\
  \beta_{0} +  \beta_{t}t_{j}+ \beta_{z}Z_{i,j} + \beta_{k} 
   & k\ge1 
  \end{cases}
\end{equation}

where  $t_{j} = (j-1)$,  $Z_{i,j}$ is a patient-level covariate (e.g. a binary indicator variable representing whether a patient characteristic is present or absent) and therefore $\beta_t$ is a time trend effect, $\beta_z$ is the patient covariate effect and $\beta_k$ is treatment's $k$ main effect. 
We shall assume that $Z_{i,j}\sim Bern(q_j)$ 
and define $\text{Expit}(u) = \frac{exp(u)}{1+exp(u)}$. Furthermore, we shall assume that the global null hypothesis is true, meaning $H_{0,k}$ holds for $k = 1,\dots,K$, or equivalently $\beta_{1} = \dots =\beta_{k} =0$.
Patients with 
$Z_{i.}=1$  will have   success rate when allocated to arm $k$ 
equal to 
$\text{Expit}(\beta_{0} + \beta_{t}t_{j}+\beta_{z})$ while patients with a negative value $Z_{i.}=0$  will have a success rate of
$\text{Expit}(\beta_{0}+\beta_{t}t_{j})$.  

If the covariate variable $Z$ is unobservable then when analysing the data, response rates will in effect be marginalised over $Z$ as follows:
\begin{eqnarray}
Pr(Y_{i,j,k}=1|a_{i,j,k}=1) &=& \sum^{1}_{f=0}Pr(Y_{i,j,k}=1|Z_{i,j}=f,a_{i,j,k}=1)Pr(Z_{i,j}=f) \nonumber \\
                       &=& \text{Expit}(\beta_{0} + \beta_{t}t_{j})(1-q_{j}) +   \text{Expit}(\beta_{0} + \beta_{t}t_{j}+ \beta_{z} )q_{j}
\end{eqnarray}
\noindent
Assuming that equal numbers of patients are recruited at each of $J$ stages then the mean response rate in arm $k$ will be
\begin{equation}\label{average}
Pr(Y_{i,.,k}=1|a_{i,.,k}=1) = \frac{1}{J}\sum^{J}_{j=1}Pr(Y_{i,j,k}=1|a_{i,j,k}=1)
\end{equation}
The inclusion of $t_j$ and $Z_{i,j}$ 
allow us not only to introduce time trends of different magnitude but also to describe two distinct scenarios that are likely to
be a concern in modern clinical trials: \emph{changes in the standard of care} (Scenario (i))  - or changes in the effectiveness of the control treatment, and  \emph{patient drift} (Scenario (ii)) - or changes in the baseline characteristics of patients. Under model \eqref{model1} we shall consider that a case of Scenario (i) occurs if $\beta_t\not=0$ while $\beta_z=0$ 
whereas an instance of Scenario (ii)  happens if $\beta_z\not =0$ while $\beta_t=0$ 
and $q_j$ evolves over $j$.

In this section we consider the global null hypothesis by setting $\beta_{k} = 0$ for all $k$  for both scenarios.
In \autoref{Sec:3.3} and \autoref{Sec:4}  we consider extensions of these scenarios where $\beta_k>0$ for some $k>1$. Specifically, we consider 
alternative hypotheses of the form $H_{1,k}: p_k- p_0=\Delta p>0$ for some $k>1$ with the treatment effect $\Delta p$  defined as  $\Delta p=Pr(Y_{i,.,k}=1|a_{i,.,k}=1) -Pr(Y_{i,.,k}=1|a_{i,.,0}=1)$.

\subsection{\normalsize RAR procedures considered}\label{sec:2.1}

Many variants of RAR have been proposed in the literature.  However, different RAR procedures often perform similarly, because they obey the same fundamental principle. 
\emph{Myopic} procedures determine the `best' allocation probabilities for the next patient (or block of patients) according to some criteria based on the accumulated data (on both responses and allocations) up to the last treated patient. \emph{Non-myopic} procedures  consider not only current data but also all possible future allocations and responses to determine the allocation probability of every patient (or block of patients) in the trial
\citep[see chapter 1 in][]{hubook}. Furthermore, RAR procedures can be considered to be \emph{patient benefit-oriented} if they are defined with the goal of maximising the exposure to a best arm (when it exists). Additionally,   RAR procedures can also be defined with the goal of attaining a certain level of statistical power to detect a relevant treatment effect, thus being \emph{power-oriented}.
RAR rules that score highly in terms of patient benefit generally have lower power. 

Thus, in order to illustrate these four types of rules we  focus on the following RAR rules: `Thompson Sampling' \emph{(Myopic-Patient benfit oriented)},  `Minimise failures given power'  \emph{(Myopic-Power oriented)}, the `Forward Looking Gittins Index rule' \emph{(Non-Myopic-patient benefit oriented)}, and its controlled version, the  `Controlled Forward Looking Gittins Index rule'\emph{(Non-Myopic-Power oriented)}.  A short summary of these approaches is now given, for a more detailed description see \citep{villarflgi}.

\begin{enumerate}
\item[(a)] \emph{`Thompson Sampling'} (TS):  \citep{thompson1933likelihood} was the first to recommend  allocating patients to treatment arms based on their posterior probability of having the largest response rate.
%
%

        We shall compute the TS allocation probabilities 
using a simple Monte-Carlo approximation. Moreover, we shall introduce a tuning parameter $c$ defined as $\frac{(j-1) \times b}{2T}$ where $(j-1) \times b$ and $T$ are the current and maximum sample size respectively. 
        This parameter tunes the \emph{aggressiveness} of TS allocation rule based in the accumulated data so that the allocation probabilities become more skewed towards the current best arm only as more and more data accumulates. Notice that TS is essentially the only class of RAR considered in \cite{Thall2015}. 
%

        \item[(b)] {`\emph{Minimise failures given power}'} (RSIHR):   \citep{rosenberger2001optimal} proposed and studied an optimal allocation ratio for two-armed trials.  It is optimal in the sense that it minimises the expected number of  failures for a fixed variance of the estimator under the alternative hypothesis that there is a positive treatment effect $\Delta p = p_1-p_0>0$.

       \begin{equation}\label{huRosenberger}
 \pi_{j,k} =\frac{ \sqrt{{p_0}}}{\sqrt{{p_0}}+\sqrt{{p_1}}},
\end{equation}

        In practice the allocation probabilities are computed by plugging in  a suitable estimate for $p_{k}$ using the data up to stage $j-1$. In our simulations we estimated the success rate parameters from the mean of its prior distribution for the first block of patients (when no data were available) and  the MLE thereafter.
        The optimal allocation ratios that extend equation \eqref{huRosenberger} for the general case in which $K>1$ do not admit a closed form, however numerical solutions can be implemented as in \citep{Tymofyeyev}.

\item[(c)]  {`\emph{Forward Looking Gittins Index} rule'} (FLGI): in  \citep{villarflgi}, we introduced a block randomised implementation of the optimal deterministic solution to the \emph{classic} multi-armed bandit problem, first derived in \citep{gijo74,gi79}. The FLGI probabilities are designed to mimic 
what a rule based on the Gittins Index (GI) would do. See Section 3 and Figure 1 in \citep{villarflgi} for a more detailed explanation of how these probabilities are defined and approximately computed via Monte-Carlo. The near optimality attained by this rule differs from the one targeted in procedure (b) in the sense that average patient outcome is nearly maximised with no constraint on the power levels that should be attained. 
Notice that before the introduction of this procedure the practical implementation of non-myopic RAR rules was severely hindered by computational issues, particularly in a multi-armed scenario.

\item[(d)]  `\emph{Controlled FLGI}' (CFLGI):
 In addition to the rule 
described in  (c),  for the multi-armed case (i.e. $K>1$) we  consider a 
group allocation rule 
 which, similarly to the  procedure proposed in \citep{trippa2012}, protects the allocation to the control treatment so  it
 never goes below $1/(K+1)$ (i.e. its fixed equal allocation probability) during the trial.
\end{enumerate}

\subsection{Simulation results}\label{sec:2.2}
In this section we present the results of various simulation studies which show, for instances of scenarios (i) and (ii) (described in detail below), the degree to which the type I error rate can be inflated for different RAR rules relative to a CR design. As is the usual case when comparing RAR procedures we consider measures of efficiency (or variability) and ethical performance, assessing which ones of them (if any) provide a better compromise between these two goals \citep{flournoy2013graphical}. We therefore also compute the expected number of patients assigned to the best treatment $(p^*)$ and expected number of patient successes  (ENS). However, under the global null considered in this section ENS and $p^*$ are identical for all designs and therefore we do not report them here. In Sections 
where we consider scenarios  under various alternative hypotheses 
we report patient benefit measures as well as power. Specifically, we report $p^*$ and the increment in the expected patient benefit that the RAR rule considered attains over a CR design, i.e. $\Delta\text{ENS}=\text{ENS}_{RAR}-\text{ENS}_{CR}$.

For each scenario a total of $5000$ trials were simulated under the global null and the same global null was tested. 
We used
z-statistics for testing with RAR rules (a), (b) and (d)  (when asymptotic normality can be assumed) and an adjusted Fisher's exact test for bandit-based  rule (c). 
The adjustment for the bandit rules chooses the cutoff value
to achieve a 5\% type-I error rate (as in \citep{villarsts}). For multi-armed trials, we use the Bonferroni correction
method to account for multiple testing and therefore
ensure that the family-wise error rate is less than
or equal to 5\%.
In all simulations  
and for all RAR rules we assumed uniform priors on all arms' success rates before treating the first block of patients. 

\subsubsection{Scenario (i): Changes in the standard of care}

The first case we consider is that of  a  linear upward trend in the outcome probability of the control arm. This could be the case of a novel surgery technique that has recently become the standard of care but it requires a prolonged initial training
period for the majority of surgeons to
become proficient in these complex procedures until ``failure'' is eliminated or reduced to a minimum constant rate. 
In terms of the model described
in equation \eqref{model1} this corresponds to varying $\beta_{t}$ with all else fixed.

\begin{figure}[h!]
\centerline{\includegraphics[trim = 0mm 70mm 0mm 70mm, clip, width=.85\textwidth, height=0.330\textheight]{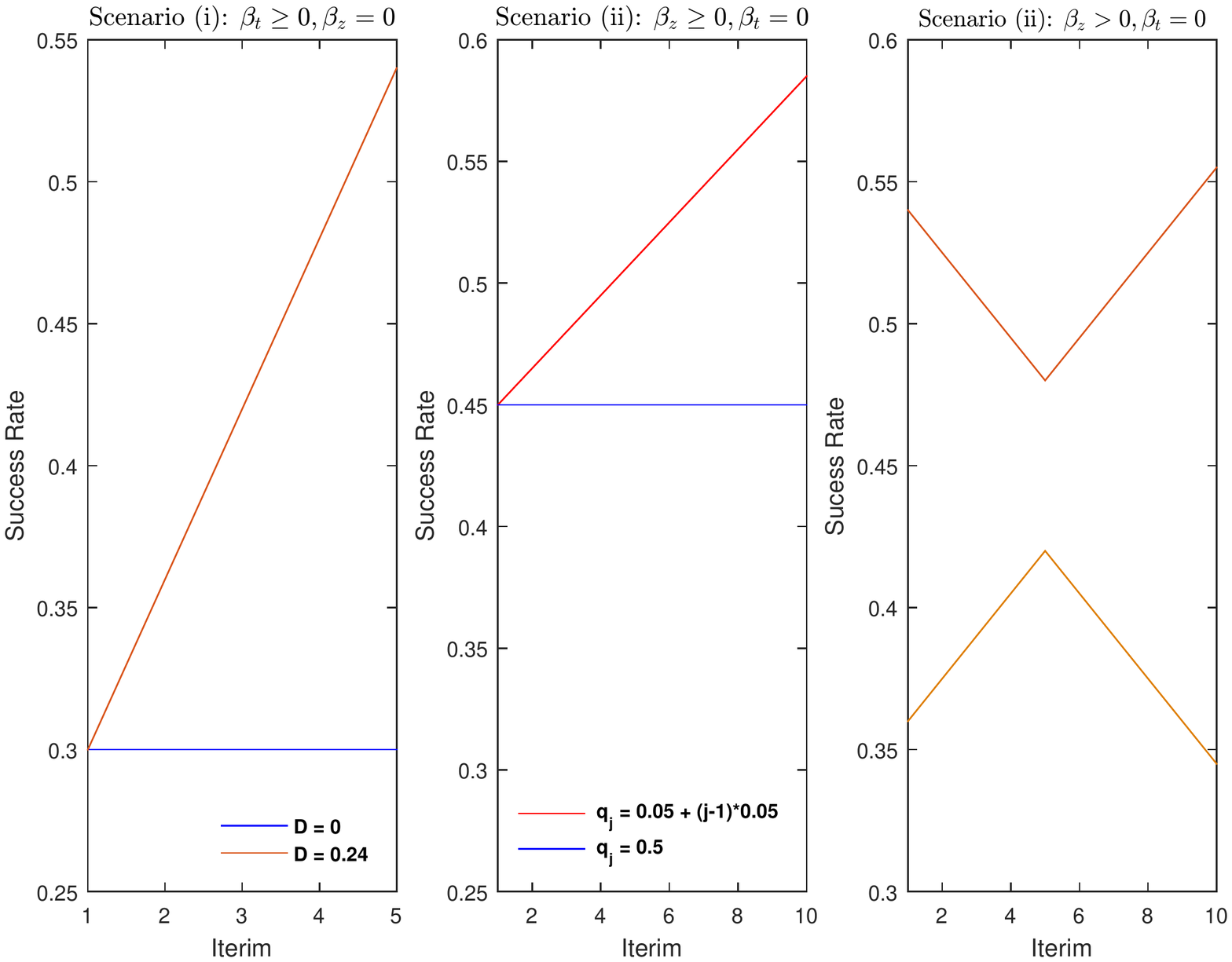}}
\caption{\small  The per block success rate under different time trend assumptions plotted over time. Left plot corresponds to scenario (i) (Changes in standard of care) and middle and right plot correspond to different cases of scenario (ii) (Patient drift).}
\label{fig:drift1}
\end{figure}

 Specifically, we  let $\beta_t$ take a value such that the overall time trend within the trial  $$D=\left[Pr(Y_{i,j,.}=1|t_{j}=J-1)\right]-\left[Pr(Y_{i,j,.}=1|t_{j}=0)\right],$$ varies in $D=\{0,0.01,0.02,0.04,0.08,0.16,0.24\}$. 
%
\autoref{fig:drift1} (left) shows the corresponding evolution of the per block success rate of every arm over time across the scenario in which $J=5$ and for the cases of:  no drift ($D=0$, dark blue) and the strongest drift considered ($D = 0.24$, dark red). 

\begin{figure}[h!]
\vspace{-6mm}
\centerline{\includegraphics[trim = 0mm 70mm 0mm 50mm, clip,width=.75\textwidth, height=0.58\textheight]{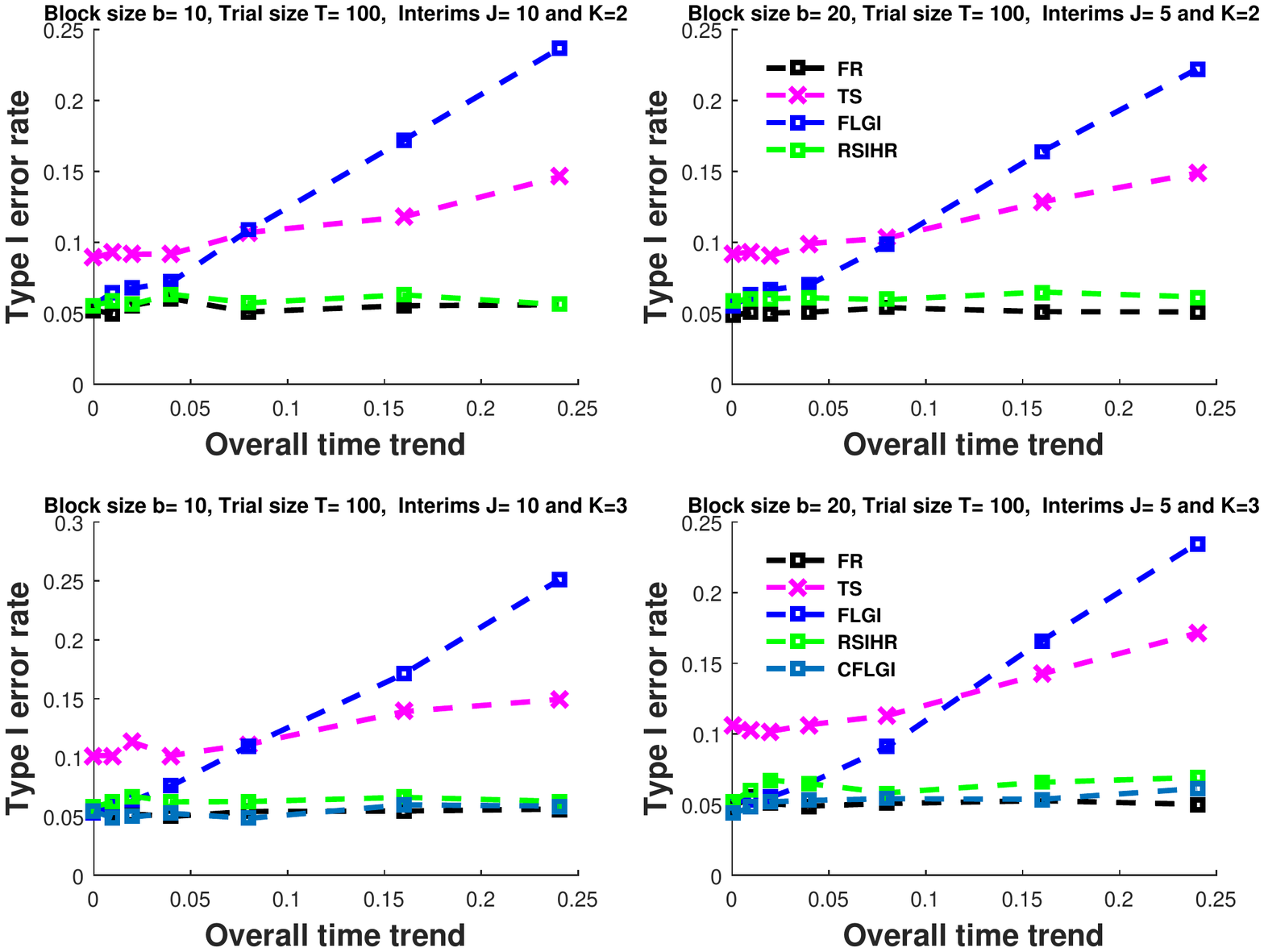}}
\caption{\small  The type I error rate for Scenario (i) (changes in the standard of care) under different linear time trends assumptions and different RAR rules.   
           }\label{fig:drift2}
               \vspace{-3mm}
\end{figure}

\autoref{fig:drift2}  summarises the simulation results. The top row plots show the results for the two-armed trials (i.e. $K=1$) and the bottom row plots show the results for $K=2$. In both cases the trial size was $T=100$. The value of the sample size $T$ might be interpreted as the maximum possible sample size (i.e. including a very large proportion of the patient population) in the context of a rare disease setting. The plots in the left column assume a block size of $10$, and the plots in the right column assume a block size of $20$.
The initial success rate was assumed to be equal to $0.3$ (which corresponds to $\beta_0\approx-0.8473$)  for all the arms considered.

Under the assumption of no time trend (i.e. $\beta_t=D=0$) the test statistics used preserve the 
type I error rate for all designs in all cases considered, except for Thompson Sampling for which the false positive rate is somewhat inflated (as pointed out in \cite{Thall2015}).
For CR the type I error rate is preserved even when a time trend is present and regardless of the block size, number of arms and the trend's magnitude.

The error rates for some of the RAR rules  (FLGI and TS) are substantial when overall time trends are of $0.08$ and more. 
This is because these   rules are \emph{patient benefit oriented}, 
i.e. they skew allocation towards an arm based on data more considerably and/or earlier on in the trial. 
 On the other hand, the RSIHR procedure, being a \emph{power-oriented}  rule, remains practically  unaffected by temporal trends in terms of type I error inflation. This is a very important difference in performance that distinguishes both RAR procedures in terms of being considerable differently affected by the same level of time trend.

Multi-arm allocation rules that protect allocation to the control arm, like the CFLGI, are also unaffected by type I error inflation, even for large drifts. Generally, the type I error inflation suffered by the other RAR rules seems to be slightly larger for the three-armed case than for the two-armed case.

\subsubsection{Scenario (ii): Patient drift}

For this case we imagine a simplistic instance in which patients are classified into two groups according to their prognosis.  This occurs if, for example, $Z_{i,j}$  in model \eqref{model1} represents the presence or absence of a biomarker in patient $i$ at stage $j$, where $Z_{i,j} = 1 $ denotes a biomarker positive patient and $Z_{i,j} = 0$ denotes  biomarker negative patient. Alternatively, $Z_{i,j}$ can capture any other patient feature. It could, for example, represent if a patient is a smoker and previously received the control arm as it would be the relevant covariate in the BATTLE-1 trial.
Moreover, we let the recruitment rates of these two types of patients, i.e. $q_j$, vary as the trial progresses to induce the desired drift in the mix of patients over time. 
We will model this situation by letting $\beta_z >0$ in \eqref{model1} whilst holding all else fixed. 
We  start by assuming that  $Z$ is  unobserved.  In \autoref{Sec:4.1} we explore the case where $Z$ is measured and can be adjusted for.

The middle and right-hand side plots in \autoref{fig:drift1}  show the evolution of $Pr(Y_{i,.,k}=1|a_{i,.,k}=1)$ under differing patterns of \emph{patient drift} over the course of the trial. 
The middle plot  describes the case in which there is a linear trend  in the average success rates of all arms created by the \emph{patient drift} whereas the plot to the right  considers the case of a more complex temporal evolution with the average success rates going up and then down or vice-versa. In both cases a trial of size $T=200$ with $J=10$ and therefore $b=20$ was considered. The success rates for all arms for the biomarker negative patients was $E[Y_{i,j,.}|Z_{i,j}=0] = 0.3$  (so that $\beta_0 \approx-0.8473$). For the biomarker positive group $E[Y_{i,j,.} = 1|Z_{i,j}=1] = 0.6$ (such that $\beta_z \approx1.2528$). 

\autoref{tab:1} summarises the simulation results for the cases depicted in \autoref{fig:drift1} (middle and right). 
The results show that  the  RAR procedures most affected by type I error inflation are those that are  \emph{patient-benefit} oriented (FLGI and TS). Type I error inflation is high only for the moderately large 
recruitment rate evolution assumed. As before, the  \emph{power-oriented} rules (RSIHR and CFLGI) have their type I error rates preserved in all cases. This further supports the argument that not all RAR procedures are equally affected by the presence of the same temporal trend. 


\begin{table}[h!]
\begin{center}
\begin{tabular}{|c|c  c|c c||c|c  c|c c}
\hline
Rule &\multicolumn{2}{|c|}{K=2}& \multicolumn{2}{|c||}{K=3} &Rule &\multicolumn{2}{|c|}{K=2}& \multicolumn{2}{|c|}{K=3}\\
\hline
 & \multicolumn{2}{|c|}{$\alpha$}     & \multicolumn{2}{|c||}{$\alpha$} & & \multicolumn{2}{|c|}{$\alpha$}     & \multicolumn{2}{|c|}{$\alpha$}\\
  \hline
 \multicolumn{5}{|c||}{$q_j=0.5+ (j-1) \times 0.05$} & \multicolumn{5}{|c|}{$q_j=0.5+(j-1)\times0.005$}\\
 \hline
 CR & \multicolumn{2}{|c|}{0.0530}  & \multicolumn{2}{|c||}{0.0504} &CR & \multicolumn{2}{|c|}{0.0478}    & \multicolumn{2}{|c|}{0.0482}\\
 TS & \multicolumn{2}{|c|}{0.1224}  & \multicolumn{2}{|c||}{0.1222}& TS & \multicolumn{2}{|c|}{0.0854}  &\multicolumn{2}{|c|}{0.0800}\\
 RSIHR &  \multicolumn{2}{|c|}{0.0558} & \multicolumn{2}{|c||}{0.0536} & RSIHR &  \multicolumn{2}{|c|}{0.0492} &\multicolumn{2}{|c|}{0.0482}\\
 FLGI& \multicolumn{2}{|c|}{0.1472} &\multicolumn{2}{|c||}{0.1726} & FLGI& \multicolumn{2}{|c|}{0.0552}& \multicolumn{2}{|c|}{0.0664 }\\
 CFLGI& \multicolumn{2}{|c|}{-}& \multicolumn{2}{|c||}{ 0.0476}& CFLGI& \multicolumn{2}{|c|}{-}& \multicolumn{2}{|c|}{0.0434}\\
  \hline
 \multicolumn{5}{|c||}{$q_j=0.5+(j-1)\times 0.05/5$} & \multicolumn{5}{|c|}{$q_j=0.5+(j-1)\times 0.05/2$}\\
 \hline
 CR & \multicolumn{2}{|c|}{0.0534}     & \multicolumn{2}{|c||}{0.0484} &CR & \multicolumn{2}{|c|}{0.0536}    & \multicolumn{2}{|c|}{0.0470}\\
 TS & \multicolumn{2}{|c|}{0.0920}  & \multicolumn{2}{|c||}{0.0888}& TS & \multicolumn{2}{|c|}{0.0914}  &\multicolumn{2}{|c|}{0.1050}\\
 RSIHR &  \multicolumn{2}{|c|}{0.0498} & \multicolumn{2}{|c||}{0.0532} & RSIHR &  \multicolumn{2}{|c|}{0.0578} &\multicolumn{2}{|c|}{ 0.0496}\\
 FLGI& \multicolumn{2}{|c|}{0.0644}&\multicolumn{2}{|c||}{0.0714} & FLGI& \multicolumn{2}{|c|}{0.0858}& \multicolumn{2}{|c|}{0.1002}\\
 CFLGI& \multicolumn{2}{|c|}{-}& \multicolumn{2}{|c||}{ 0.0452 }& CFLGI& \multicolumn{2}{|c|}{-}& \multicolumn{2}{|c|}{ 0.0450 }\\
 \hline
 \multicolumn{5}{|c||}{{$q_j=0.2+ (j-1) \times0.05$} (*)} & \multicolumn{5}{|c|}{\small{$q_j=0.8-(j-1)\times 0.05 $} (*)}\\
\multicolumn{5}{|c||}{{$q_j= 0.35 -(j-6)\times 0.05 $} (**)} &\multicolumn{5}{|c|}{\small{$q_j=0.65+ (j-6)\times 0.05$} (**)}\\
 \hline
 CR & \multicolumn{2}{|c|}{0.0354}     & \multicolumn{2}{|c||}{0.0312} &CR &\multicolumn{2}{|c|}{0.0408}     & \multicolumn{2}{|c|}{0.0286} \\
 TS & \multicolumn{2}{|c|}{0.0772}  & \multicolumn{2}{|c||}{0.0830}& TS & \multicolumn{2}{|c|}{0.0728}     & \multicolumn{2}{|c|}{0.0716} \\
 RSIHR &  \multicolumn{2}{|c|}{0.0548} & \multicolumn{2}{|c||}{0.0492} & RSIHR &  \multicolumn{2}{|c|}{0.0498} & \multicolumn{2}{|c|}{0.0456}\\
 FLGI& \multicolumn{2}{|c|}{0.0716}&\multicolumn{2}{|c||}{0.0684} & FLGI& \multicolumn{2}{|c|}{0.0474}&\multicolumn{2}{|c|}{0.0512}\\
 CFLGI& \multicolumn{2}{|c|}{-}& \multicolumn{2}{|c||}{0.0378}& CFLGI& \multicolumn{2}{|c|}{-}& \multicolumn{2}{|c|}{0.0428}\\
 \hline
 \multicolumn{5}{|c||}{\small{$q_j=0.5 $}} & \multicolumn{5}{|c|}{$q_j=0.7$}\\
 \hline
 CR &  \multicolumn{2}{|c|}{0.0542}    & \multicolumn{2}{|c||}{ 0.0522 }&CR & \multicolumn{2}{|c|}{0.0492}    & \multicolumn{2}{|c|}{0.0482} \\
 TS & \multicolumn{2}{|c|}{0.0728}  & \multicolumn{2}{|c||}{ 0.0766}& TS & \multicolumn{2}{|c|}{0.0720}  &\multicolumn{2}{|c|}{0.0756}\\
 RSIHR &  \multicolumn{2}{|c|}{0.0684}  &\multicolumn{2}{|c||}{ 0.0524} & RSIHR &  \multicolumn{2}{|c|}{0.0534} &\multicolumn{2}{|c|}{0.0514}\\
 FLGI& \multicolumn{2}{|c|}{0.0500}&\multicolumn{2}{|c||}{0.0560} & FLGI& \multicolumn{2}{|c|}{0.0520}& \multicolumn{2}{|c|}{ 0.0570}\\
 CFLGI& \multicolumn{2}{|c|}{-}& \multicolumn{2}{|c||}{ 0.0410}& CFLGI& \multicolumn{2}{|c|}{-}& \multicolumn{2}{|c|}{0.0402}\\
 \hline
\end{tabular}
\end{center}
\caption{
\small
 The type I error rate under different group recruitment rates assumptions under scenario (ii) with $\beta_z\approx 1.2528$ 
and success rates evolving as shown in Figure 1 (middle)  and Figure 1 (right). (*) $j<6$ and (**) $j\ge6$.
}\label{tab:1}
\end{table}

\section{Testing procedures and RAR designs robust to patient drift}\label{Sec:3}

In this section we describe
 a hypothesis testing procedure for RAR rules in a two-armed trial context and a RAR design for multi-armed trials that preserves type I error rates in the presence of an unknown time trend.

\subsection{Two-armed trials: randomization test and the FLGI}\label{Sec:3.1}

The type I error inflation shown in scenarios (i) and (ii) for some of the RAR rules is caused by the fact that the test statistics used assume  every possible sequence of treatment allocations (i.e., every possible trial  realisation) is equally likely. For instance, this is the case for the adjusted
Fisher's exact test used for the FLGI in the previous sections.
 This assumption is not true in general as certain allocation sequences will be highly unlikely or even impossible for some RAR procedures. This is particularly well illustrated in the case of the FLGI rule where   it is possible for one of the arms to be effectively `selected' within the trial, since the probability of assigning a patient to the other arm from that point onwards is zero.

In this section we show the results of developing and computing a test statistic, introduced in \citep{simon2011}, based on the distribution of the  assignments induced by the FLGI under the null hypothesis.
In their paper, the authors show that 
using a cut-off value from the  distribution of the test statistic  
generated by the RAR rule under the null hypothesis, and conditional on the vector of observed outcomes, ensures the control of the type I error rate (see Theorem 1 in \citep{simon2011}). Their result applies to any RAR rule and any time trend in a two-armed trial, most importantly, its implementation  does not require any knowledge or explicit modelling of the trend. In this paper we have chosen to implement it for the FLGI rule as this is the most recently proposed RAR procedure of the ones considered.

 However,  computation of the null distribution can be challenging under realistic trial scenarios as it requires the complete enumeration of all trial histories and it is infeasible for response adaptive rules that are deterministic as e.g. the GI rule is. 
  Therefore, there is a need to find ways of computing such a randomization test efficiently for the sake of its practical implementation as well as evaluating its effect on power, which might differ across different rules.

 We implement  a randomisation test for the FLGI rule that is based on a Monte-Carlo approximation of the exact randomisation test. More precisely, our approach does the following: for a given trial history $\mathbf{y}=(\mathbf{y}_1, \mathbf{y}_2, \dots, \mathbf{y}_J)$, where $\mathbf{y}_j$ a vector of the $b$ observed outcomes at stage $j$, we simulate $M$ trials under the FLGI allocation rule. 
The FLGI allocation ratios are updated after each block using  the allocation variables $a_{i,j.k}$ randomly generated under the FLGI rule by Monte-Carlo and the observed outcome data up to that point (i.e. $(\mathbf{y}_{1},...,\mathbf{y}_{j})$). For each simulated trial  we compute the value of the test statistic to assemble  an empirical distribution of the test statistic under the null.  We can then compare the test statistic observed in the original trial to the empirical distribution, rejecting the  null hypothesis at level $\alpha$ if it is more extreme than its $\alpha$ percentile for a one-sided test (or than its $\alpha/2$ or $1-\alpha/2$ percentile for a two-sided test). Finally, we repeat this procedure for another $Nr$ trial history replicates and report the average type I error rate achieved as well as the averages of the other ethical performance measures considered.

The results from $Nr=5000$ replicates of using the approximate randomisation test for the case of scenario (i)  displayed in \autoref{fig:drift2} (top-right) are shown in \autoref{tab:2}. For each trial replicate the approximate randomisation-based test  used $M=500$ simulated trials to construct the  empirical distribution function. The values of $Nr$ and $M$ are the same as those used by the simulations in \cite{simon2011}. From \autoref{tab:2}, we see that the type I error rate is preserved at its $5\%$ level even when the patient drift is severe. We also report $p^{*}$ and $\Delta ENS$, as defined in Section 2. 

\begin{table}[h!]
\begin{center}
\begin{tabular}{c|cccc|}
$\text{Rule}$& $\alpha \text{ (s.e.)}$  & $p^* \text{ (s.e.)}$  &$\Delta\text{ENS}$  & $D$  \\
\hline
\hline
\emph{$FLGI_{20}$} &0.0445 (0.21) & 0.501 (0.21) & $\phantom{-}0.19$ &      0\\
\emph{$FLGI_{20}$} & 0.0480 (0.21) &0.506 (0.22) & $-0.17$  & 0.08        \\
\emph{$FLGI_{20}$} & 0.0449 (0.20) &   0.494 (0.23) & $\phantom{-}0.02$  & 0.16 \\
\emph{$FLGI_{20}$} &  0.0445 (0.21) &  0.499 (0.24) & $\phantom{-}0.23$       &  0.24\\
\hline
\end{tabular}
\end{center}
\vspace{5mm}
\caption{
\small
The type I error rate for the approximate randomisation test from $5000$ replicates of  
a 2-arm trial of size $T=100$ using a FLGI with  block size $b=20$ ($J=5$) and  under the case of Scenario (i) depicted in \autoref{fig:drift2} (top-right plot). 
}\label{tab:2}
\end{table}
\subsection{Multi-armed trials: protecting allocation to control}\label{Sec:3.2}

As shown in the simulation results reported in \autoref{Sec:2} the RAR rules that include a protection of the allocation to the control treatment (specifically, the CFLGI) preserve the type I error rate. Matching the number of patients allocated to control to that allocated to the best performing arm also ensures that standard analysis methods can be applied to detect significant differences between the two groups with a high power.

Therefore, if the design of the multi-arm trial incorporates protection of the control allocation there appears to be no need to implement a testing procedure that is specifically designed to be robust to type I error inflation. 

\subsection{ Protecting against time trends and its effect on Power}\label{Sec:3.3}

Preserving the type I error rate is an important requirement for a clinical trial design. However, the learning goal of a trial also requires that, if a best experimental treatment exists, then the design should also have a high power to detect it. In this section, we therefore assess the  power of the approximate randomisation test (for the FLGI) and the standard test (for the CFLGI).

We first explore an extension of an instance of scenario (i) 
in which we assume there is a treatment effect of $0.4$ (where $p_0=0.3$ and $p_1=0.7$, therefore $\beta_1\approx 1.6946$) which is maintained even in the cases where we also assume a positive time trend in the standard of care. The trial is of size $T=150$ with $J=5$ stages (so that $b=30$). Under this design the assumed treatment effect is detected with approximately $80\%$ power by the FLGI rule if there is no time trend and the adjusted Fisher's test is used. If a traditional CR design is used then the power attained is $99\%$, but the proportion of patients allocated to each arm is fixed at 1/2. 

\autoref{tab:3} shows the power  to reject the null hypothesis as the overall time trend increases from 0 to $0.24$ (i.e. for $\beta_t \in \{0, \dots 
,0.27\}$ while $\beta_z=0$) for a treatment effect of $0.4$ (i.e. for $\beta_1\approx 1.6946$).  We denote by $(1-\beta_F)$ the power level attained by the adjusted Fisher's test and by $(1-\beta_{RT})$  the power level when using the approximate randomisation  test.

These results show that the power of the randomisation test is considerably reduced compared to that obtained using Fisher's exact test.  
However, the patient benefit properties of the FLGI over the CR design are preserved in all the scenarios. The improvement in patient response of the FLGI design over CR is around 15\% regardless of the drift assumption. 


Next, we consider the multi-arm case by assessing the effect on power on the RAR rules that protect allocation to the control arm.
 In order to do so under different trend assumptions we extend a case of scenario (i). 
 We assume then that there is a treatment arm that has an additional benefit over the other two arms of $0.275$ (where $p_0=p_2=0.300$ and $p_1=0.575$) which is maintained even in the cases we assume a positive time trend in the success rate of the standard of care. Regardless of the trend assumption,  a traditional CR design has a mean $p^{*}$ value of 1/3 by design and detects a treatment effect of such a magnitude with approximately $80\%$ power.

  \autoref{tab:4} shows the power levels and other operating characteristics for the designs considered. 
  The CR design performs as predicted in terms of power, $p^{*}$ and ENS.
  The power of the CFLGI is unaffected except for a small increase when the trend is very high. 
  This approach attains an improvement over CR on $p^*$ and ENS for every trend magnitude assumption considered (the improvement in ENS goes from  15.81\% to 13.44\%  in the case of the largest assumed trend). 

\autoref{tab:4} also includes the results for the  FLGI for comparison. Power levels are increased when there is a positive time trend, as the ones assumed in this case, compared to the case when there is no time trend. Such an increment is caused by the temporal upward trend which for the FLGI rule causes an overestimation of the treatment effect. 
This table also illustrates how the GI-based methods introduce a larger variability in the resulting patient allocation per arm (a point raised by \cite{Thall2015} about TS). However, the table also shows that this increased variability is twice as much for the FLGI than for the CFLFI. As well, with respect to the probability of this allocation imbalance being in the wrong direction (i.e. towards inferior arms) GI-based rule perform extremely well as this only occurs in less than 4\% of all replicates.

\begin{table}[h!]
\begin{center}
\begin{tabular}{c |c c c c c|}
\textbf{} & 
\multicolumn{1}{c}{\textbf{$(1-\beta)_F$}}(s.e.)  & $(1-\beta)_{RT}$&  $p^*$ (s.e.)  &$\Delta$ENS (s.e.) & $D$ \\
\hline
\hline
\emph{$FLGI_{30}$} &  0.8086 (0.39) &   0.6057 (0.48) & 0.871 (0.09) & 22.04 & 0\\
\emph{$FLGI_{30}$} &  0.8972 (0.30) &   0.6080 (0.49) & 0.881 (0.04) & 24.01 &   0.08       \\
\emph{$FLGI_{30}$} &  0.9524 (0.21) &   0.6021 (0.50) & 0.878 (0.05) &  23.99 &   0.16 \\
\emph{$FLGI_{30}$} &  0.9802 (0.14) &   0.5851 (0.48) & 0.882 (0.03) & 23.73 &   0.24\\
\hline
\end{tabular}
\end{center}
\vspace{5mm}
\caption{
\small Power for the approximate randomisation test from $5000$ replicates of  
a 2-arm trial of size $T=150$ using a FLGI with  block size $b=30$ ($J=5$) under a case of Scenario (i)  with a treatment effect of 0.40. 
}\label{tab:3}
 \end{table}

\begin{table}[h!]
\begin{center}
\begin{tabular}{|c |c c c c|}
\hline
\textbf{} & 
\multicolumn{1}{c}{\textbf{$(1-\beta)$}}(s.e.)  & $p^*$ (s.e.)  &$ENS$ (s.e.) & $D$ \\
\hline
\emph{$CR_{20}$} &  0.8164 (0.39) & 0.333 (0.04) & 58.76 (5.92) & 0       \\
\emph{$CR_{20}$} &  0.8096 (0.39) & 0.333 (0.04) & 59.68 (6.11) & 0.01  \\
\emph{$CR_{20}$} &  0.8122 (0.39) & 0.333 (0.04) & 60.26 (6.09) & 0.02  \\
\emph{$CR_{20}$} &  0.8010 (0.40) & 0.334 (0.04) & 61.70 (6.05) & 0.04     \\
\emph{$CR_{20}$} &  0.8020 (0.40) & 0.335 (0.04) & 64.87 (6.07) & 0.08  \\
\emph{$CR_{20}$} &  0.7990 (0.40) & 0.334 (0.04) & 70.80 (6.00) & 0.16     \\
\hline
\emph{$CFLGI_{20}$} & 0.8782 (0.33) & 0.552 (0.07) & 68.04 (6.83) & 0  \\
\emph{$CFLGI_{20}$} & 0.8744 (0.33) & 0.554 (0.07) & 69.01 (6.74) & 0.01\\
\emph{$CFLGI_{20}$} & 0.8752 (0.33) & 0.553 (0.08) & 69.87 (6.76) & 0.02  \\
\emph{$CFLGI_{20}$} & 0.8760 (0.33) & 0.551 (0.08) & 70.99 (6.76) & 0.04 \\
\emph{$CFLGI_{20}$} & 0.8844 (0.32) & 0.553 (0.08) & 74.23 (6.67) & 0.08\\
\emph{$CFLGI_{20}$} & {0.8920} (0.31) & 0.552 (0.08) & 80.26 (6.83) & {0.16} \\
\hline
\emph{$FLGI_{20}$} & 0.6612 (0.47) & 0.755 (0.12) & 76.82 (7.92) & 0\\
\emph{$FLGI_{20}$} & 0.6802 (0.47) & 0.756 (0.12) & 77.65 (7.84) & 0.01\\
\emph{$FLGI_{20}$} &  0.6984 (0.46) & 0.758 (0.12) & 78.31 (7.89) & 0.02 \\
\emph{$FLGI_{20}$} &  {0.7228} (0.45) & 0.758 (0.12) & 80.07 (7.85) & {0.04} \\
\emph{$FLGI_{20}$} & {0.7896} (0.41) & 0.762 (0.12) &  83.10 (7.83) & {0.08}   \\
\emph{$FLGI_{20}$} & {0.8696} (0.34) & 0.761 (0.12) &  89.14 (7.91) & {0.16}  \\
\hline
\end{tabular}
\end{center}
\caption{
\small Power of CR, CFLGI and FLGI in
$5000$ replicas of  
a 3-arm trial of size $T=100$  with  block size $b=20$ ($J=5$) under a case of scenario (i) with a treatment effect of $0.275$ for arm 1
}\label{tab:4}
\end{table}

\section{Adjusting the model for a time trend}\label{Sec:4}

In this section we illustrate the extent to which adjusting for covariates can help to reduce type I error inflation and affect power. 
This section also discusses the problems that can be encountered when doing this after having used a RAR procedure  and how to address them. 

\subsection{Two-armed trials}\label{Sec:4.1}
We first study covariate-adjustment under instances of scenario (i).  We consider a two-armed trial of size $T=100$ with $J=5$ and $b=20$. We shall  focus on the most extreme case considered in \autoref{fig:drift1} in which the overall time trend was $D=0.24$ (or $\beta_t\approx0.2719$). The initial success rates of both arms were set to $0.3$ (i.e. $\beta_0\approx -0.8473$). 

Parts (I) and (III) in \autoref{tab:5} show the results for the estimation of the models' parameters using standard maximum likelihood estimation, when the (logistic) model is correctly specified. These results indicate, perhaps unsurprisingly, that for both designs the treatment effect is found to be significant in less than 5\% of the 5000 trials, which suggests that  by including a correctly modelled time trend,  type I error inflation is avoided. However, we note that there is a strong deflation in the type I error rate of the FLGI design. This occurs because  the testing procedure used in  this section does not include an adjustment similar to the one used with Fisher's exact test in \autoref{Sec:2} and \autoref{Sec:3}. When we look at the mean estimated coefficient for the time trend we note that CR only slightly underestimates it, having a 40\% power to detect it as significantly different from 0. The FLGI design results in a larger underestimation of the time trend coefficient.  
This underestimation is consistent with that observed in \citep{villarsts} for reasons clarified in \citep{bowden2015unbiased}. The power to detect a significant time trend for the FLGI is more than halved compared to CR. Since its estimate is negatively correlated with that of the time trend coefficient,
the baseline effect $\beta_0$ is also overestimated in both designs, but more severely for the FLGI. 

Another consequence of the under-estimation of the time trend is that complete or quasi-complete separation is more likely to occur \citep[See][]{albertandanderson}. This happens  for the FLGI for example when all the observations of one of the  arms are failures (and few in number) and this arm is therefore dropped early from the trial (i.e. its allocation probability goes to $0$ and never goes above $0$ again within the trial).

\renewcommand{\arraystretch}{1.4}
\begin{table}[h!]
\begin{center}
\begin{tabular}{|c |c c c|}
\hline
\multicolumn{4}{|c|}{(I) GLM fitting without correction for CR}\\
\hline
\ \ \ \   \phantom{$\hat{\beta_0}$} \ \ \  & \ \ \ \ \  $\Ex{\left(\hat{\beta_i}\right)}$  \ \ \ & \ \  \ \  $\Ex\left(\text{MSE}\right)$ \ \ \ & $\Ex\left(p_{value} <0.05\right) $\\
 \hline
 $\hat{\beta_0}$& -0.8684& 0.1992  &     0.5174\\
$\hat{\beta_t}$& \phantom{-}0.2610 & 0.0243  &     0.4018   \\
$\hat{\beta_1}$& \phantom{-}0.0070 & 0.1900 &     0.0544  \\
    \hline
    \multicolumn{4}{|c|}{(II) GLM fitting with correction for CR}\\
    \hline
$\hat{\beta_0}$ &-0.8370 & 0.1838  &      0.5224\\
$\hat{\beta_t}$        &\phantom{-}0.2509 &0.0227    &   0.4012\\
$\hat{\beta_1}$   &\phantom{-}0.0067 & 0.1775     &  0.0534\\
\hline
\multicolumn{4}{|c|}{(III) GLM fitting without correction for FLGI}\\
\hline
$\hat{\beta_0}$&-1.4465 &8.9957 &       0.4110\\
$\hat{\beta_t}$    &   \phantom{-}0.1898 &0.0307  &     0.1844  \\
$\hat{\beta_1}$ &  \phantom{-}0.0038 &18.2440   &     0.0142\\
    \hline
    \multicolumn{4}{|c|}{(IV) GLM fitting with correction for FLGI}\\
    \hline
$\hat{\beta_0}$ &-0.9307 & 0.3947  &       0.4670\\
$\hat{\beta_t}$        &\phantom{-}0.1825 &0.0301    &   0.1858\\
$\hat{\beta_1}$   &\phantom{-}0.0048 & 0.7993     &  0.0456\\
\hline
\end{tabular}
\end{center}
\vspace{1mm}
\label{tab:glmscenario1a}
\caption{GLM estimated through MLE with and without Firth correction for $T=100$, $J=5$, $b=20$ in a case of scenario (i) 
with $D=0.24$. Results for 5000 trials. True values were assumed to be $\beta_0\approx -0.8473$, $\beta_t\approx0.2719$ and $\beta_z=\beta_1=0$. }\label{tab:5}
\end{table}


When this problem occurs in a trial realization, 
the maximum likelihood estimates are highly unstable and 
will not be well defined. This can be observed in the MSE value for $\hat{\beta_0}$ in \autoref{tab:5} (III) for the FLGI. In order to address this, we applied Firth's penalized likelihood approach \citep{firth}  which is a method for dealing with issues of separability, small sample sizes, and bias of the parameter estimates (using the R package ``logistf''). In \autoref{tab:5} parts (II) and (IV) results of deploying the Firth correction are displayed. These results show an improvement in the estimation of the baseline effect when using the FLGI design: the MSE value is significantly reduced and the average estimate of $\beta_0$ is closer to its true value (though it is still overestimated). For the CR design there is also an improvement. Also, note that the type I error deflation has also been almost fully corrected by the Firth's adjustment in the FLGI design.

To assess the effect on statistical power in \autoref{tab:6} we replicate the estimation procedure for the case studied in the $3^{rd}$ row of \autoref{tab:3} in which  we let the treatment effect of arm 1  be positive (i.e.  $\beta_1\approx1.6946$) 
while the overall drift assumed corresponds with $D=0.16$ (or $\beta_t\approx0.1840$ and $\beta_z=0$). 
The  initial success rate in the control arm  was equal to $0.3$ (i.e. $\beta_0\approx -0.8473$). 
Because complete (or quasi-complete) separation affected the FLGI rule in all the scenarios considered here, \autoref{tab:6}
and the following tables  only display the results using Firth's correction. 

As expected the power of a CR design displayed in \autoref{tab:6} coincides with the value reported in \autoref{Sec:3.3}, which using both procedures (i.e. adjusting for covariates or hypothesis testing) yields an average value of 99\%. The power value of the FLGI design when fitting the GLM model is close to the value reported in \autoref{Sec:3.3} for the case of no time trend (i.e. $\approx$ 80\%). The difference is caused by the adjustment in Fisher's exact test done in that section which raises power by slightly overcorrecting for the deflation of the type I error rate of the standard Fisher's test.

Our results suggest that correctly modelling a time trend and adjusting for separation via Firth correction can safeguard the validity of trial analyses using RAR. That is, by maintaining correct type I error rates and delivering a level of statistical power similar to that obtainable when no trend is present.




\renewcommand{\arraystretch}{1.4}
\begin{table}[h!]
\begin{center}
\begin{tabular}{|c |c c c|}
\hline
    \multicolumn{4}{|c|}{(II) GLM fitting with correction for CR}\\
    \hline
\ \ \ \   \phantom{$\hat{\beta_0}$} \ \ \  & \ \ \ \ \  $\Ex{\left(\hat{\beta_i}\right)}$  \ \ \ & \ \  \ \  $\Ex\left(\text{MSE}\right)$ \ \ \ & $\Ex\left(p_{value} <0.05\right) $\\
\hline
$\hat{\beta_0}$ & -0.8951 &0.1413  &   0.7194\\
$\hat{\beta_t}$   & \phantom{-}0.1985& 0.0175   &    0.3262\\
$\hat{\beta_1}$   &\phantom{-}1.7831 &0.1488    &   0.9994\\
\hline
    \multicolumn{4}{|c|}{(IV) GLM fitting with correction for FLGI}\\
    \hline
$\hat{\beta_0}$ & -0.8832&0.3192    &   0.3364\\
$\hat{\beta_t}$ & \phantom{-}0.2408 &0.0291    &   0.3062\\
$\hat{\beta_1}$ & \phantom{-}1.6917 &0.4313    &   0.7394\\
\hline
\end{tabular}
\end{center}
\vspace{1mm}
\label{tab:glmscenario1a}
\caption{GLM estimated through MLE with Firth correction for $T=150$, $J=5$, $b=30$ 
in a case of Scenario (i) with $D=0.16$ and $\beta_1=1.6946$. Results for 5000 trials. True values were assumed to be $\beta_0\approx -0.8473$, $\beta_t\approx0.1840$, $\beta_1=1.6946$ and $\beta_z=0$ }\label{tab:6}
\end{table}

\subsection{Multi-armed trials}\label{Sec:4.1}

In this section we consider the case of multi-armed trials and  an instance of Scenario (ii) or \emph{patient drift}.    Also, we shall remove the assumption that the patient covariate or biomarker is unobservable, and allow for the availability of this information before analysing and estimating the corresponding model in \eqref{model1}.

First, we study the case of scenario (ii) in which the proportion of biomarker positive patients evolves  as $q_j=0.5+ (j-1) \times 0.05$ for $j=1, \dots,10$ (See \autoref{fig:drift1}, middle). We simulated $5000$ three-armed trials of size $T=200$ with $J=10$ and $b=20$. The differential effect of being biomarker positive was assumed to be of $0.3$, which corresponds with  $\beta_z\approx1.2528$.  
The initial success rates of all the arms  for the biomarker negatives was equal to $0.3$ (i.e. $\beta_0\approx -0.8473$). 

\begin{table}[h!]
\begin{center}
\begin{tabular}{|c |c c c|}
   \hline
    \multicolumn{4}{|c|}{(II) GLM fitting with correction for CR}\\
    \hline
\ \ \ \   \phantom{$\hat{\beta_0}$} \ \ \  & \ \ \ \ \  $\Ex{\left(\hat{\beta_i}\right)}$  \ \ \ & \ \  \ \  $\Ex\left(\text{MSE}\right)$ \ \ \ & $\Ex\left(p_{value} <0.05\right) $\\
\hline
$\hat{\beta_0}$ & -0.8527 & 0.1307 &      0.6778  \\
$\hat{\beta_z}$   &\phantom{-}1.2597 & 0.1142 &      0.9758   \\
$\hat{\beta_1}$   &-0.0084& 0.1305 &      0.0458 \\
$\hat{\beta_2}$   &-0.0029& 0.1304 &      0.0486  \\
   \hline
    \multicolumn{4}{|c|}{(IV) GLM fitting with correction for FLGI}\\
    \hline
$\hat{\beta_0}$ & -0.8771 & 0.1724  &    0.6740  \\
$\hat{\beta_z}$ &  \phantom{-}1.2471 & 0.1169   &   0.9702    \\
$\hat{\beta_1}$ &   \phantom{-}0.0114 & 0.2228   &    0.0598  \\
$\hat{\beta_2}$ &  -0.0097 & 0.2246  &    0.0620 \\
    \hline
    \multicolumn{4}{|c|}{(VI) GLM fitting with correction for CFLGI}\\
    \hline
$\hat{\beta_0}$ & -0.8455 & 0.1338   &    0.6632\\
$\hat{\beta_z}$ & \phantom{-}1.2505 &0.1200    &  0.9686   \\
$\hat{\beta_1}$ &  -0.0226&  0.1471  &    0.0558\\
$\hat{\beta_2}$ &  -0.0196 &0.1477    &   0.0492  \\
\hline
\end{tabular}
\end{center}
\vspace{1mm}
\label{tab:glmscenario1a}
\caption{GLM estimated through MLE with Firth correction for $T=200$, $J=10$, $b=20$, $K=3$ 
in a case of scenario (ii) in which $q_j=[0.5:0.05:0.95]$. Results for 5000 trials. True values were assumed to be $\beta_0\approx -0.8473$, $\beta_z\approx 1.2528$ and $\beta_t=\beta_1=\beta_2=0$}\label{tab:7}
\end{table}

 \autoref{tab:7} displays the results of the CR, FLGI and CFLGI designs under the null hypothesis. These results suggest that all  designs attain the same power to detect the biomarker effect (as the adaptation is not done using this information, all designs have similar numbers of patients with a positive and a negative biomarker status). 
More importantly, all of the designs correct for the type I error inflation if the patient covariate is completely observable and incorporated into the explicative model.

 In  \autoref{tab:8} we examine the effect on power by replicating the previously described scenario but allowing for the experimental arm 1 to have an effect for all patients' types of $0.2$ (i.e. $\beta_1=0.8473$). These results show how the power to detect the treatment effect in arm 1 with a FLGI design is almost halved compared to that attained by a traditional CR design. Yet, the CFLGI improves on the power level of the CR design by approximately 15\%. Also note that the type I error rate for arm 2 appears to be deflated for the FLGI and CFLGI designs. 

These results suggest that fitting a model that includes a time
trend after having used a RAR rule can protect against the type I error
inflation caused by \emph{patient drift}  as long as the patient-covariate information is observable and available to adjust for.
However, the power level attained by covariate adjustment is considerably less than that attained by a design that protects the allocation to the control arm.

Furthermore, these results fail to illustrate the learning-earning trade-off that characterises the choice between a CR and a RAR procedure and the reasons why the FLGI  could be desirable to use from a patient benefit perspective  (despite the power loss and the type I error inflation potential). The traditional CR design, which maximises learning about all arms, yields an average number of successfully treated patients (or ENS) of $116.83$ when $p^*$ 
remains fixed by design at 1/3. 
The FLGI design, on the other hand, is almost optimal from a patient benefit perspective, achieving an ENS value of $135.21$,  15.73\% higher than with CR, and it achieves this by skewing $p^*$ 
to $0.7783$. Finally, the CFLGI is a compromise between the two opposing goals that improves on the power levels attained by a CR design and also on its corresponding ENS value (though is below the value that could be attained with the unconstrained FLGI rule) by attaining an ENS vale of $126.32$, 8.12\% more than with a CR design,  and a $p^*$ 
of $0.5618$.

\section{Discussion}\label{Sec:5}

Over the past 65 years, RCTs have become the gold standard approach for evaluating treatments in human populations. Their inherent ability to protect against sources of bias is undoubtedly one of their most attractive features, and is also the reason that many are unwilling to recommend the use of RAR rules, feeling that this would be a
 ``step in the wrong direction'' \citep{simon77}. Recently, \cite{Thall2015} have suggested that a severe type I error inflation could occur if RAR is used under the presence of an unaccounted for time trend. However, there is also a strong interest in the medical community to use RAR procedures in contexts where there are several arms in a rare disease setting.


In this paper we have assessed by simulation the level of type I error inflation of several 
RAR procedures by creating scenarios that are likely to be a concern in modern clinical trials that have a long duration. 
Our results suggest that the magnitude of the temporal trend necessary to seriously inflate the type I error of the \emph{patient benefit oriented}  RAR rules needs to be of an important magnitude (i.e. change larger than a 25\% in its outcome probability) to be a source of concern. This supports the conclusion of \citep{karrison2003group} 
in a group sequential design context. However, we also conclude that the trend magnitude does not seem to significantly affect some RAR rules. Specifically not those that are power oriented such as the CFLGI rule \citep{villarflgi} or the `\emph{Minimise failures given power}' rule  \citep{rosenberger2001optimal}. This suggests that when giving drawbacks to the use of RAR in real trials one must be careful not to include all RAR rules in the same class, as they have markedly different performances in the same situation.

In addition, we have recommended two different procedures that can be used in an RAR design to protect for type I error inflation. For two-armed trials, the use of a randomisation test (instead of traditional tests) preserves type I error under any type of unknown temporal trend.
The cost of this may be 
a computational burden and an reduction in 
statistical power (although most patients are still allocated to a superior arm when it exists). 
This particular feature highlights the need to develop 
computationally feasible 
testing procedures that are specifically tailored to the behaviour of a given RAR rule. For example, as pointed out by \citep{villarsts}, bandit based rules such as the FLGI are extremely successful at identifying the truly best treatment but, as a direct result, often  cannot subsequently declare its effect `significant' using standard testing methods. 

For multi-armed clinical trials protecting allocation to the control group (the recommended procedure) preserves the type I error while yielding a power increase with respect to a traditional CR design. 
 However, despite rules such as the CFLGI being more robust to a time trend effect, they also offer a reduced patient benefit in the case there is a superior treatment, when compared to the \emph{patient benefit oriented}  RAR rules such as the 
FLGI.

Finally, we also assessed  adjustment for a time trend  both as an alternative protection procedure against type I error inflation and to highlight estimation problems that can be encountered when an RAR rule is implemented. Our conclusion is that adjustment can  alleviate the type I error inflation of RAR rules (if the trend is correctly specified 
and the associated covariates are measured and available
). However, for the multi-armed case this strategy attains a lower power than simply protecting the allocation to the control arm. Furthermore, the technical problem of separation also complicates estimation after the \emph{patient benefit oriented}  RAR rules have been implemented and severely impacts the power to detect a trend compared to an CR design. 

 Further research is needed to assess the potential size of time trends through careful re-analysis of previous trial data (as \citep{karrison2003group} do with data from \citep{kalish1987impact}).
However, there are some RAR rules that, both for the two-armed and the multi-armed case, remain largely unaffected in all the cases we have considered. Of course, these rules offer increased patient benefit properties  when compared to a traditional CR design but reduced when compared to the \emph{patient benefit oriented}  RAR rules. We believe this to be one of the most important contributions of the present work, highlighting the importance of carefully choosing a RAR procedure when patient drift is a concern.

Another area of future work is to explore the combination of randomisation tests with stopping rules in a group sequential context, specifically for the FLGI.
Additionally, techniques for the efficient computation of approximate randomisation tests for the FLGI could be studied, similar to
 those explored in
\citep{rosenbergerannals}. 



\begin{table}[h!]
\begin{center}
\begin{tabular}{|c |c c c|}
\hline
    \multicolumn{4}{|c|}{(II) GLM fitting with correction for CR}\\
    \hline
\ \ \ \   \phantom{$\hat{\beta_0}$} \ \ \  & \ \ \ \ \  $\Ex{\left(\hat{\beta_i}\right)}$  \ \ \ & \ \  \ \  $\Ex\left(\text{MSE}\right)$ \ \ \ & $\Ex\left(p_{value} <0.05\right) $\\
\hline
$\hat{\beta_0}$ & -0.8816 & 0.1324   &    0.7078\\
$\hat{\beta_z}$   & \phantom{-}1.3006 &  0.1239 &      0.9726 \\
$\hat{\beta_1}$   & \phantom{-}0.9355 & 0.1549  &     0.6954\\
$\hat{\beta_2}$   & -0.0032 &0.1356  &     0.0516\\
\hline
    \multicolumn{4}{|c|}{(IV) GLM fitting with correction for FLGI}\\
    \hline
$\hat{\beta_0}$ & -1.1635 &0.5391    &   0.3994\\
$\hat{\beta_z}$ &  \phantom{-}1.3492 & 0.1300   &    0.9762   \\
$\hat{\beta_1}$ & \phantom{-}1.1300 &0.5845    &   0.3672  \\
$\hat{\beta_2}$ & \phantom{-}0.0041 &0.7740     &  0.0246 \\
\hline
    \multicolumn{4}{|c|}{(VI) GLM fitting with correction for CFLGI}\\
    \hline
$\hat{\beta_0}$ & -0.8861 &0.1378   &    0.6966\\
$\hat{\beta_z}$ &  \phantom{-}1.3127 &0.1243   &    0.9718  \\
$\hat{\beta_1}$ &  \phantom{-}0.8862 &0.2077   &    0.7718\\
$\hat{\beta_2}$ &  -0.2487& 0.5027  &     0.0288 \\
\hline
\end{tabular}
\end{center}
\vspace{1mm}
\label{tab:glmscenario1a}
\caption{GLM estimated through MLE with Firth correction for $T=200$, $J=10$, $b=20$, $K=3$ 
in a case of scenario (ii) in which $q_j=[0.5:0.05:0.95]$. 
Results for 5000 trials. True values were assumed to be $\beta_0\approx -0.8473$, $\beta_z\approx 1.2528$ and $\beta_1=0.8473$, $\beta_t=\beta_2=0$}\label{tab:8}
\end{table}

\section*{Acknowledgements}
This work was funded by the UK Medical Research Council (grant numbers G0800860,  MR/J004979/1, and  MR/N501906/1) and the Biometrika Trust.





\bibliographystyle{unsrtnat}

\end{document}